\documentclass[aps,prb,twocolumn,showpacs,tightenlines,amsmath]{revtex4}
 \usepackage{graphics}
 \usepackage{epsfig}
   \begin{document} 
   \title{\Large \bf{
             Time-resolved noise of adiabatic quantum pumps
                    }
         }
\author{
M. Moskalets$^{1,2}$
and
M. B\"uttiker$^1$
}
\affiliation{
     $^1$D\'epartement de Physique Th\'eorique, Universit\'e de Gen\`eve,
     CH-1211 Gen\`eve 4, Switzerland\\
     $^2$Department of Metal and Semiconductor Physics,\\
     National Technical University "Kharkiv Polytechnic Institute",
     61002 Kharkiv, Ukraine\\
}
\date\today
   \begin{abstract}
We investigate quantum-statistical correlation properties of a periodically 
driven mesoscopic scatterer on a time-scale shorter than the period of a drive.
In this limit the intrinsic quantum fluctuations in the system of fermions
are the main source of a noise.
Nevertheless the effect of a slow periodic drive is clearly visible in a two-time 
current-current correlation function as a specific periodic in time modulation.
In the limit of a strong drive such a modulation can change the sign
of a current correlation function.
\end{abstract}
\maketitle
\small

\section{Introduction}
\label{intro}

Experimental detection \cite{SMCG99} of a dc current generated
by the mesoscopic sample in response to local slow periodic driving
revived interest in the adiabatic quantum pump effect and 
stimulated increased experimental and theoretical efforts.
While there are still only a few experiments 
\cite{SMCG99,HLWB01,WPMU03,DMH03,VDM04}
the theoretical literature is large and we can refer here only to a few representative works
\cite{
BTP94,Brouwer98,PB03,AEGS04,Vavilov05,SGKF06}.

One of the possible applications of an adiabatic pump consists in using it
as a source of correlated particle flows. \cite{SamuelssonB04,BTT05,MBen05} 
Adiabatic driving excites particles only a little 
and thus avoids strong dephasing due to inelastic relaxation.
However for quantum information processing not only phase coherence but in addition 
the correlations between emitted particles are crucial, since only non-classical
\cite{Schrodinger35,EPR35,Bohr35,Bell65}
correlations can be used, see, e.g., Ref.~\onlinecite{BD00}.

In mesoscopic systems the zero-frequency noise 
is usually used to characterize the correlations between the particles. 
\cite{Buttiker90,SB05,Beenakker05}
In a time-dependent set-up this quantity provides information averaged over a pump period 
which characterizes, in particular, how regularly the pump emits particles. 
This follows from the observation that the regime
of quantized pumping \cite{HN91,AA98}
(when exactly the same number of particles is emitted for each pumping cycle) 
is characterized by a vanishing \cite{AK00,MM01,AEGS01}
zero-frequency noise power.
Thus a pump emitting particles regularly does not produce a zero-frequency noise.

The properties of a pump evolve in time. 
That in turn makes the properties of emitted particles dependent 
on the time moment when they actually leave the pump. 
In general the particles emitted to different leads, leave the pump
at different time moments. 
Consider, for instance, the processes taking place in a pump in a quantized pumping regime.
Typically this is a resonant transmission structure subject to large amplitude driving.
\cite{WWG00,KAEW04,DR05}
In this case the quantization of the pumped charge is due to 
well separated quantum levels inside the structure 
which are filled and emptied during the pumping cycle. 
When the quantum level is filled then one particle (we ignore spin) 
is transferred from the reservoir to the pump. 
Alternatively this process can be view as emission of a hole.
Further, when the level is emptied then one particle is transferred from the pump
to the reservoir. 
To achieve a directed transport between the reservoirs, say, $\alpha$ and $\beta$
it is necessary to have the pump strongly coupled to the reservoir $\alpha$
(and decoupled from other reservoirs) during the former process
and to the reservoir $\beta$ during the latter one, or vice versa. \cite{KAEW04}
The coupling conditions change during the pumping cycle. 
As a result the current pulses (corresponding to emitted particles)
at different leads occur at different time moments. 

Therefore, to get a more detailed description of correlation properties of emitted particles
it is necessary to investigate the noise properties of a pump 
on a time-scale shorter than the pump period.
Our aim is to calculate and explore a two-time current-current correlator
for a periodically driven mesoscopic scatterer.
We expect that this quantity will be useful to access additional information
concerning correlations between emitted particles. 

We use the approach presented in 
Ref.~\onlinecite{MBstrong02}
where, on the basis of a scattering matrix approach to ac transport 
in phase coherent mesoscopic system \cite{BTP94,Buttiker92},
the general Floquet scattering matrix approach, see, e.g., Ref.~\onlinecite{PA04},
was adapted to describe the low frequency limit of a cyclic scatterer.

The paper is organized as follows.
In Sec.\ref{ccgen} we describe the approach used and the approximations employed. 
In Sec.\ref{2tccf} we calculate the time-dependent current generated by the pump and investigate the current correlation function in the time domain. 
To illustrate the general approach in Sec.\ref{2tscs} we consider a resonant transmission pump.
We conclude in Sec.\ref{concl}.

\section{General expressions}
\label{ccgen}

To investigate the basic physical phenomena underlying the quantum pump effect
we omit unnecessary complications and consider the following model: 
We suppose
(i) the pump to be connected to $N_{r}$ macroscopic particle reservoirs via single-channel ballistic leads. 
The reservoirs are in equilibrium and they are not affected by the presence of the pump. 
(ii) the electrons to be spinless noninteracting particles which maintain phase coherence propagating through the pump from one reservoir to another reservoir,
and 
(iii) the characteristic energy-scales involved to be considerably smaller 
than the Fermi energy $\mu$ of electrons.

     \subsection{Floquet scattering matrix approach}

The  model presented above can be effectively described within the scattering approach.
\cite{Buttiker92}
In each lead $\alpha = 1, 2, \cdots, N_r$ we introduce two kinds of (second quantization) operators in energy representation, 
$\hat {a}_{\alpha}, \hat {a}_{\alpha}^{\dagger}$ and $\hat {b}_{\alpha}, \hat {b}_{\alpha}^{\dagger}$.
They correspond to incoming to the scatterer and out-going off the scatterer electrons, respectively. 
If the properties of a scatterer evolve in time periodically (the period is ${\cal T}=2\pi/\Omega$) then $\hat {b}$-operators are related to $\hat {a}$-operators via the Floquet scattering matrix 
${\mathbf S}_F$: \cite{MBstrong02}
\begin{equation}
\label{Eq6}
\begin{array}{c}
\hat {b}_{\alpha}(E) = \sum\limits_{E_{n}>0}\sum\limits_{\beta=1}^{N_r} {S}_{F,\alpha\beta}(E,E_{n}) \hat {a}_{\beta}(E_{n});
\ \\
\hat {b}_{\alpha}^{\dagger}(E) = \sum\limits_{E_{n}>0}\sum\limits_{\beta=1}^{N_r} 
{S}_{F,\alpha\beta}^{\dagger}(E_{n},E)\hat {a}_{\beta}^{\dagger}(E_{n}).
\end{array}
\end{equation}
Here $E_n=E+n\hbar\Omega$, where $n$ is the number of energy quanta which the electron absorbs ($n<0$) or emits ($n>0$) during interaction with an oscillating scatterer.

The current conservation forces ${\mathbf S}_{F}$ to be a unitary matrix:
\begin{equation}
\label{Eq5}
\begin{array}{c}
\sum\limits_{E_{n}>0} {\mathbf S}_{F}^{\dagger}(E_{m},E_{n})
{\mathbf S}_{F}(E_{n},E) = {\mathbf I}\delta_{m,0}; \\
\ \\
\sum\limits_{E_{n}>0} {\mathbf S}_{F}(E_{m},E_{n})
{\mathbf S}_{F}^{\dagger}(E_{n},E) = {\mathbf I}\delta_{m,0}.
\end{array}
\end{equation}
Here ${\mathbf I}$ is a unit $N_{r}\times N_{r}$ matrix.

In terms of operators for incoming and out-going particles 
the current $I_{\alpha}$ in lead $\alpha$ reads as follows:\cite{Buttiker92}
\begin{subequations}
\label{Eq9}
\begin{equation}
\label{Eq9A}
I_{\alpha}(\omega) = e\int\limits_{0}^{\infty}dE
 \langle
\hat b_{\alpha}^{\dagger}(E)\hat b_{\alpha}(E+\hbar\omega)
- \hat a_{\alpha}^{\dagger}(E)\hat a_{\alpha}(E+\hbar\omega)
\rangle,
\end{equation}
\begin{equation}
\label{Eq9B}
I_{\alpha}(t) = \frac{e}{h}
\int\limits_{0}^{\infty}\int\limits_{0}^{\infty}
dEdE^{\prime} e^{\frac{i}{\hbar}(E-E^{\prime})t}
\langle
\hat b_{\alpha}^{\dagger}(E)\hat b_{\alpha}(E^{\prime})
- \hat a_{\alpha}^{\dagger}(E)\hat a_{\alpha}(E^{\prime})
\rangle,
\end{equation}
\end{subequations}
where $\langle\cdots\rangle$ denotes a quantum-statistical average.

Another quantity we investigate is the correlation function of currents
(see, e.g., Ref.~\onlinecite{BB00}):
\begin{subequations}
\label{Eq10}
\begin{equation}
\label{Eq10A}
{ P}_{\alpha\beta}(\omega,\omega^{\prime}) = \frac{1}{2}
\langle 
\Delta\hat I_{\alpha}(\omega)
\Delta\hat I_{\beta}(\omega^{\prime})
+ \Delta\hat I_{\beta}(\omega^{\prime}) \Delta\hat I_{\alpha}(\omega)
\rangle,
\end{equation}
\begin{equation}
\label{Eq10B}
{ P}_{\alpha\beta}(t_{1},t_{2}) = \frac{1}{2}
\langle 
\Delta\hat I_{\alpha}(t_{1})
\Delta\hat I_{\beta}(t_{2})
+ \Delta\hat I_{\beta}(t_{2}) \Delta\hat I_{\alpha}(t_{1})
\rangle.
\end{equation}
\end{subequations}
Here $\Delta\hat I = \hat I - \langle \hat I \rangle$ is a current fluctuation operator. 

The quantities in time and in frequency representation are related to each other
via the Fourier transformation:
\begin{equation}
\label{Eq11}
\begin{array}{c}
X(t) = \frac{1}{2\pi}\int\limits_{-\infty}^{\infty} d\omega 
e^{-i\omega t}X(\omega), \\
\ \\
X(\omega) = \int\limits_{-\infty}^{\infty} dt e^{i\omega t}X(t).
\end{array}
\end{equation}
In the equations for correlation functions such a transformation
is applied to each argument.

Notice that throughout the paper we deal with two kinds of frequencies
which we suppose to be small compared to the Fermi energy,
\begin{equation}
\label{Eq12}
\hbar\Omega, \hbar\omega \ll \mu.
\end{equation}
The first is a fixed pump frequency $\Omega$. 
The second is a variable measurement frequency $\omega$.
The former defines an energy shift in the Floquet scattering matrix elements
and thus relates to the scattering properties of the driven conductor of interest, while the latter gives us spectral contents of a measured quantity. 

     \subsection{Current and current correlation function}

Substituting Eq.(\ref{Eq6}) into Eqs.(\ref{Eq9A}) we get the current:
\begin{subequations}
\label{Eq13}
\begin{equation}
\label{Eq13A}
\begin{array}{c}
I_{\alpha}(\omega) = \sum\limits_{l=-\infty}^{\infty}
2\pi\delta(\omega - l\Omega) {\cal J}_{\alpha, l},\\ 
\ \\
{\cal J}_{\alpha,l} = 
\frac{e}{h}\int\limits_{0}^{\infty}dE
\sum\limits_{\beta}
\sum\limits_{n} 
\big\{f_{\beta}(E_n) - f_{\alpha}(E) \big\} \\ 
\times
S_{F,\alpha\beta}^{*}(E,E_n)S_{F,\alpha\beta}(E_l,E_n),
\end{array}
\end{equation}
{\rm where 
$$
f_{\alpha}(E)=\langle a^\dagger_\alpha(E)a_\alpha(E)\rangle=\frac{1}{1+\exp\frac{E-\mu_\alpha}{k_BT_{\alpha}}}
$$ 
is the Fermi distribution function with $T_\alpha$ and $\mu_{\alpha}$ being the temperature and the chemical potential of reservoir $\alpha$; $k_B$ is the Boltzmann constant.
For the correlation function, Eq.(\ref{Eq10A}), we find}
\begin{equation}
\label{Eq13B}
\begin{array}{c}
{ P}_{\alpha\beta}(\omega,\omega^{\prime}) = \sum\limits_{l=-\infty}^{\infty}
\pi \delta(\omega + \omega^{\prime} - l\Omega) 
{\cal P}_{\alpha\beta, l}(\omega,\omega^{\prime}),\\ 
\ \\
{\cal P}_{\alpha\beta, l}(\omega,\omega^{\prime}) = 
\frac{e^2}{h}\int\limits_{0}^{\infty}dE
\bigg\{ \delta_{\alpha,\beta}\delta_{l,0}
f_{\alpha\alpha}(E,E+\hbar\omega) \\
- \sum\limits_{n}f_{\alpha\alpha}(E,E+\hbar\omega) 
S_{F,\beta\alpha}^{*}(E_{n}+\hbar\omega,E+\hbar\omega)
S_{F,\beta\alpha}(E_{n+l},E) \\
- \sum\limits_{n}f_{\beta\beta}(E,E+\hbar\omega^{\prime}) 
S_{F,\alpha\beta}^{*}(E_{n}+\hbar\omega^{\prime},E+\hbar\omega^{\prime})
S_{F,\alpha\beta}(E_{n+l},E) \\
+ 
\sum\limits_{\gamma}\sum\limits_{\delta}
\sum\limits_{n}\sum\limits_{m}\sum\limits_{q}
f_{\gamma\delta}(E_n + \hbar\omega, E_{l+n-q}) \\
\times
S_{F,\alpha\gamma}(E+\hbar\omega, E_n+\hbar\omega)
S_{F,\alpha\delta}^{*}(E,E_{l+n-q}) \\
\times
S_{F,\beta\delta}(E_{l+n-m}, E_{l+n-q})
S_{F,\beta\gamma}^{*}(E_{n-m}+\hbar\omega,E_{n}+\hbar\omega)
\bigg\}.
\end{array}
\end{equation}
\end{subequations} 
Here we introduced a convenient abbreviation:
$$
f_{\alpha\beta}(E,E^{\prime}) = 
f_{\alpha}(E)[1-f_{\beta}(E^{\prime})] + 
f_{\beta}(E^{\prime})[1-f_{\alpha}(E)].
$$

First of all from Eq.(\ref{Eq13A}) it follows  that the oscillatory scatterer
generates currents having a discrete frequency spectrum 
$\omega = l\Omega$, $l=0,\pm 1, \pm 2,\dots$. 
Therefore, the current generated is periodic in time with a period of 
${\cal T} = 2\pi/\Omega$.
The harmonic with $l=0$ is a dc current being usually the quantity of interest.
In general there are two sources of dc currents.
First, we can have a potential difference between 
the reservoirs, $\mu_{\alpha}~-~\mu_{\beta}~\neq~0$,
and, second, an oscillatory scatterer generates its own dc current (a pump effect). 
The higher harmonics, $l\neq 0$, (at stationary reservoirs) are solely due to 
a non stationary scatterer. 

In contrast to the current, the noise power, ${\cal P}$, exists even if the pump
does not work (i.e., in the stationary case) and the reservoirs are at the same
conditions. 
In such a case only one term in Eq.(\ref{Eq13B}) survives,
$$
{ P}_{\alpha\beta}^{(st)}(\omega,\omega^{\prime}) = 
\pi\delta(\omega + \omega^{\prime}) 
{\cal P}_{\alpha\beta, 0}^{(st)}(\omega,\omega^{\prime}).
$$
This term is due to quantum fluctuations present in the system with
continuous spectrum, see, e.g., Ref.~\onlinecite{Gardiner}. 
They are visible in the noise power but they are not visible in the current. 

\begin{figure}[t]
  \vspace{0mm}
  \centerline{
   \epsfxsize 8cm
   \epsffile{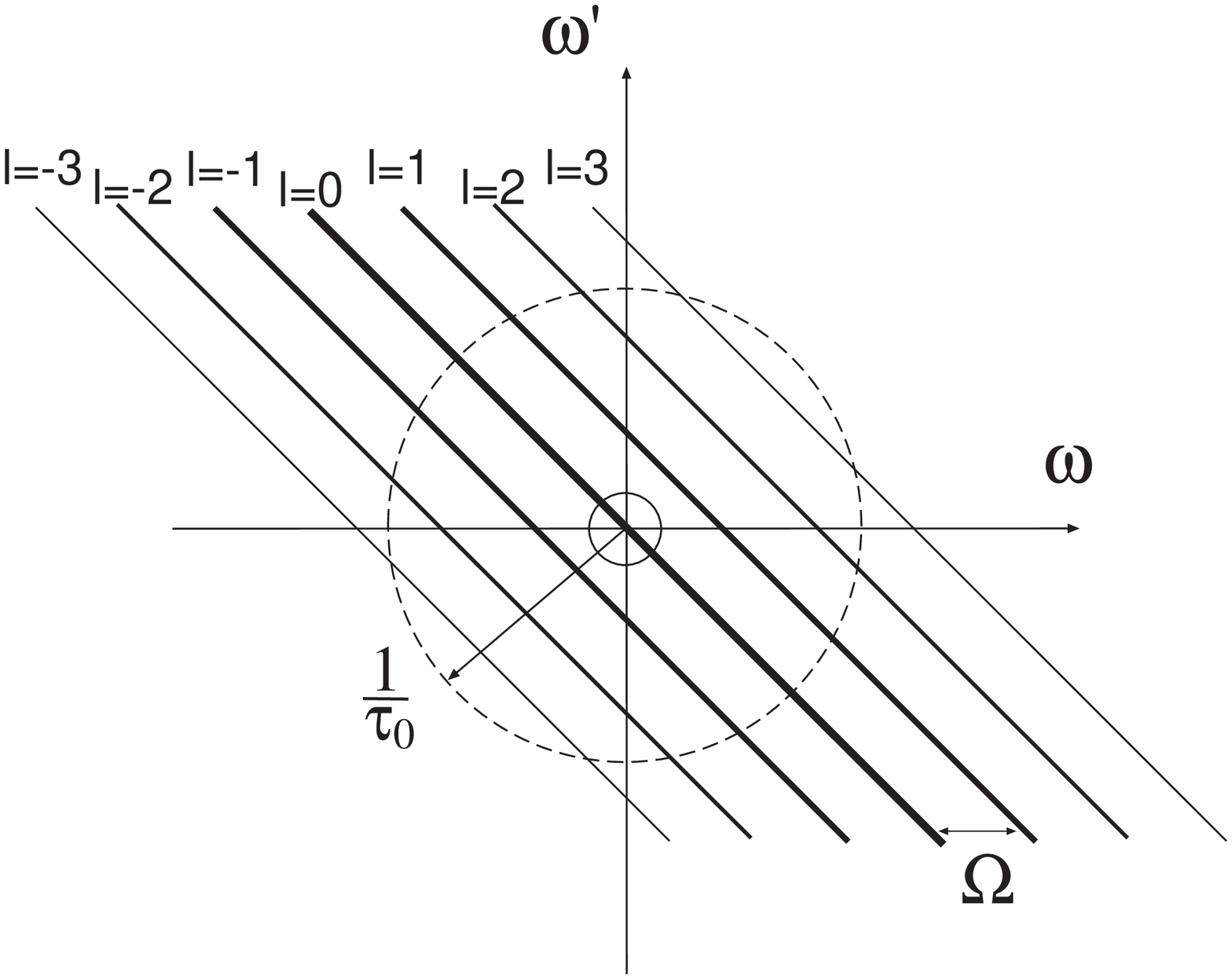}
             }
  \vspace{0mm}
  \nopagebreak
  \caption{The correlation function
${ P}_{\alpha\beta}(\omega,\omega^{\prime})$,
Eq.(\ref{Eq13B}),
for currents generated by the pump
is non-zero along the straight lines 
in the plane ($\omega$,$\omega^{\prime}$).
The lines corresponding to $l=0,\pm 1, \pm2, \pm 3$ are shown.
The width of lines shows schematically a noise intensity decreasing with
increasing number $|l|$. 
The area bounded by the dashed circle with radius $\tau_0^{-1}$
contributes to the two-time current correlation function 
${ P}_{\alpha\beta}(t_1,t_2;\tau_0)$,
Eq.(\ref{Eq30B}).
Here $\tau_0$ is a measurement time period.
The small solid circle bounds the area contributing to the zero-frequency noise power, Eq.(\ref{Eq39}).
}
\label{fig0}
\end{figure}

A working pump has a twofold effect on the measured noise, Fig.\ref{fig0}.
First, it gives rise to side lines with $l\neq 0$ in the plane
$(\omega,\omega^{\prime})$ [see, Eq.(\ref{Eq13B})],
and, second, it modifies the noise power at the main line, $l=0$. 
Therefore, one can not simply divide the measured noise into 
a quantum noise and a noise generated by the pump. 
They interfere strongly between themselves. 
The existence of a quantum noise makes it difficult to analyze the current 
correlations on a time-scale of the order of the pump period ${\cal T}$. 
Nevertheless, as we will show, the pump has a unique and 
distinguishable effect on the measured current fluctuations.

In general it is difficult to analyze Eqs.(\ref{Eq13}) analytically.
Therefore, to proceed further we make some simplifications. 
First, since we are interested in the current generated by the pump,
we consider the case when there are no currents due to other reasons.
In particular we suppose that all the reservoirs have the same potentials
and temperatures:
\begin{equation}
\label{Eq15}
\mu_{\alpha} = \mu, \quad T_{\alpha} = T, \quad \alpha = 1,\dots, N_{r}.
\end{equation}
Second, we consider the particular but important case of small driving frequencies. 
We suppose $\Omega$ to be small enough to apply an adiabatic approximation
to calculate the Floquet scattering matrix.

     \subsection{Adiabatic approximation}

In general, to calculate the Floquet scattering matrix ${\mathbf S}_{F}(E,E_n)$ 
one needs to solve the full time-dependent scattering problem.  
Here we are interested in the limit of low driving frequencies. 
In this limit we use an adiabatic approximation and express the Floquet 
scattering matrix in terms of the stationary scattering matrix 
${\mathbf S}_{0}$ which is assumed to be known. 
The adiabatic approximation is valid if the energy-scale $\hbar\Omega$ dictated 
by the modulation frequency is small compared with the energy-scale 
$\delta E$ over which the stationary scattering matrix ${\mathbf S}_{0}(E)$ 
changes significantly: \cite{MBstrong02,MBac03}
\begin{equation}
\label{Eq17}
  \hbar\Omega \ll \delta E \ll\mu.
\end{equation} 
Notice, under the condition given above the side band energy $E_n = E + n\hbar\Omega$
is positive for any reasonable $n$ and $E\sim\mu$ of interest here.
Therefore, in all the relevant equations we can safely extend summation 
over the discrete index from $-\infty$ to $+\infty$.

Let the stationary scattering matrix ${\mathbf S}_{0}$ depends on some parameters 
$p_j \in \{p\} , j = 1,2,\dots ,N_p$.
We suppose that the external driving results in a variation of these parameters, $p_j=p_j(t), \forall j$.
The matrix ${\mathbf S}_{0}(E,t)={\mathbf S}_{0}(E,\{p(t)\})$ with parameters to be fixed at time moment $t$ we call a frozen scattering matrix.
Since the driving is periodic the parameters and in turn the frozen scattering matrix are periodic in time as well.

To zero-th order in driving frequency 
the elements of the Floquet scattering matrix  ${\mathbf S}_{F}(E_n,E)$
can be approximated by the Fourier coefficients  
${\mathbf S}_{0,n}$ of the frozen scattering matrix ${\mathbf S}_{0}(E,t)$ 
as follows: \cite{MBstrong02,MBac03}
\begin{subequations}
\label{Eq18}
\begin{equation} 
\label{Eq18A} 
{\mathbf S}_{F}(E_n,E) = {\mathbf S}_{0,n}(E) + \mathcal{O}(\Omega).  
\end{equation} 
\begin{equation} 
\label{Eq18B} 
{\mathbf S}_{F}(E,E_n) = {\mathbf S}_{0,-n}(E) + \mathcal{O}(\Omega).  
\end{equation} 
\end{subequations}
Here $\mathcal{O}(\Omega)$ denotes the rest which is at  
least of the first order in $\Omega$  and which we neglect. 

The Fourier transformation used reads as follows
\begin{subequations} 
\label{Eq19} 
\begin{equation} 
\label{Eq19A} 
{\mathbf S}_{0}(E,t) = \sum\limits_{n=-\infty}^{\infty}e^{-in\omega t} 
{\mathbf S}_{0,n}(E),  
\end{equation} 
\begin{equation} 
\label{Eq19B} 
{\mathbf S}_{0,n}(E) = \int\limits_{0}^{\cal T} \frac{dt}{{\cal T}}
e^{in\omega t}  {\mathbf S}_{0}(E,t). 
\end{equation} 
\end{subequations} 
Notice, since the frozen scattering matrix is periodic in time we expand it in 
the Fourier series in contrast to a Fourier integral expansion
used in Eqs.(\ref{Eq9}) - (\ref{Eq11}).

     \subsection{Low temperature approximation}

The next simplifications we make concerns other energy-scales
in the problem of interest. 

The effect of the pump is strongly pronounced if the energy quantum 
$\hbar\Omega$ is larger than (or comparable to) the temperature. 
Therefore, taking into account Eq.(\ref{Eq17}), 
we suppose the temperature to be
less than the energy-scale $\delta E$ relevant for the scattering matrix,
\begin{equation}
\label{Eq20}
   k_BT \ll \delta E.
\end{equation} 

The pump generates currents and noise
at frequencies of order $\Omega$.
Therefore, we will consider the current
and noise at $\omega\sim \Omega$ only. 
Thus, in addition to Eqs.(\ref{Eq17}) and (\ref{Eq20}) we put
\begin{equation}
\label{Eq21}
\hbar\omega, \hbar\omega^{\prime} \ll \delta E.
\end{equation}

In such a low temperature and low frequency limit 
the scattering matrix can be treated energy independent,
${\mathbf S}_{0}(E) \equiv {\mathbf S}_{0}(\mu)$.
Then, taking into account Eq.(\ref{Eq15}),
we perform an energy integration in Eqs.(\ref{Eq13}) as follows: 
\begin{equation}
\label{Eq22}
\begin{array}{c}
\int\limits_{0}^{\infty}dE 
\big\{f_{\beta}(E_n) - f_{\alpha}(E) \big\} =  -n\hbar\Omega, \\
\ \\
\int\limits_{0}^{\infty}dE f_{\alpha\beta}(E,E+\Delta E) = 
\Delta E \coth\left(\frac{\Delta E}{2k_BT} \right).
\end{array}
\end{equation}
Substituting Eqs.(\ref{Eq18}) and (\ref{Eq22}) into Eq.(\ref{Eq13}), 
we find the current spectral density and the noise power
in the adiabatic low temperature limit as follows:
\begin{subequations}
\label{Eq23}
\begin{equation}
\label{Eq23A}
{\cal J}_{\alpha,l} = 
-i\frac{e}{2\pi} 
\left({\mathbf S}_{0}(\mu)
\frac{\partial{\mathbf S}_{0}^{\dagger}(\mu)}{\partial t} \right)_{\alpha\alpha,l},
\end{equation}
\begin{equation}
\label{Eq23B}
\begin{array}{c}
{\cal P}_{\alpha\beta, l}(\omega,\omega^{\prime}) = 
\frac{e^2}{h}\bigg\{ 
\Big(\delta_{\alpha,\beta}\delta_{l,0} 
- \big|S_{0,\beta\alpha}(\mu) \big|^2_{l}
\Big) \hbar\omega \coth\left(\frac{\hbar\omega}{2k_BT}\right) \\
- \big|S_{0,\alpha\beta}(\mu) \big|^2_{l}
\hbar\omega^{\prime} \coth\left(\frac{\hbar\omega^{\prime}}{2k_BT}\right) \\
+ \sum\limits_{q} \big|{\mathit \Sigma}_{\alpha\beta}(\mu) \big|^2_{l-q,q}
\hbar[(l-q)\Omega - \omega] 
\coth\left(\frac{\hbar[(l-q)\Omega - \omega]}{2k_BT}\right)
\bigg\}.
\end{array}
\end{equation}
\end{subequations}
Here the lower index $l$ denotes a Fourier transformed 
squared matrix element [see, Eq.(\ref{Eq19B})].
${\mathit \Sigma}_{\alpha\beta}$ is an element of a {\it two-particle} (frozen)
scattering matrix 
\begin{equation}
\label{Eq_new1}
{\mathbf \Sigma}(t_1,t_{2};E) = 
{\mathbf S}_{0}(t_{1},E) {\mathbf S}_{0}^{\dagger}(E,t_{2}),
\end{equation}
introduced in Ref.~\onlinecite{MBnoise04}.
The Fourier coefficients for this matrix entering Eq.(\ref{Eq23B}) can be expressed in terms of the single-particle scattering matrix:
\begin{equation}
\label{Eq_new2}
\big|{\mathit \Sigma}_{\alpha\beta}\big|^2_{l-q,q} = 
\sum\limits_{\gamma}\sum\limits_{\delta}
\Big(S_{0,\alpha\gamma}S_{0,\alpha\delta}^{*} \Big)_{l-q}
\Big(S_{0,\beta\gamma}^{*}S_{0,\beta\delta}\Big)_{q}.
\end{equation}

Now we use Eqs.(\ref{Eq23}) to calculate the current correlation function in the time domain.

\section{Two-time current correlation function} 
\label{2tccf}

The current correlation function 
${ P}_{\alpha\beta}(t_{1},t_{2})$  diverges at coincident times $t_{1}=t_{2}$.
That is due to quantum fluctuations having an unbounded spectrum.
To avoid such infinities we will integrate any time-dependent quantity 
over some time interval $\tau_0$. 
The additional advantage of such a procedure is that we will find a quantity which can be easily
interpolated between a time-averaged quantity (at $\tau_0\to\infty$)
and a local in time one (at $\tau_0\to 0$).
Since we are investigating the correlations on a time-scale of the order of a pump period, we put $\tau_0\sim {\cal T}$. 
Strictly speaking we will consider $\tau_0$ both larger or smaller than ${\cal T}$.
However in order for the equation (\ref{Eq23B}) to remain valid
the interval $\tau_0$ must be 
larger than $\hbar/\delta E$ [see, Eq.(\ref{Eq17})]. 

Thus, for any time-dependent measurable $X(t)$ 
we introduce a corresponding quantity $X(t;\tau_{0})$
calculated in the following way:
\begin{equation}
\label{Eq28}
X(t;\tau_{0}) = \frac{1}{\sqrt{2\pi}\tau_{0}}
\int\limits_{-\infty}^{\infty} d\tau e^{-\frac{(\tau-t)^2}{2\tau_{0}^2}} X(\tau).
\end{equation}
For the sake of computational simplicity we use the Gaussian kernel.
Using the inverse Fourier transformation, Eq.(\ref{Eq11}), 
we express $X(t;\tau_0)$
in terms of a spectral density $X(\omega)$ as follows:
\begin{equation}
\label{Eq29}
X(t;\tau_0) = \frac{1}{2\pi}
\int\limits_{-\infty}^{\infty} d\omega 
e^{-i\omega t}X(\omega)
e^{-\frac{1}{2}(\omega\tau_0)^2}.
\end{equation}
Notice, in dealing with the current correlation function 
${ P}_{\alpha\beta}(t_{1},t_{2})$,
Eq.(\ref{Eq10B}), we use for 
the first and second time arguments the Gaussian kernels
centered at different time moments $t_{1}$ and $t_{2}$, respectively.

Using Eqs.(\ref{Eq13}) and Eq.(\ref{Eq29})
we find the current $I_{\alpha}(t;\tau_0)$
and the two-time current correlation function 
${ P}_{\alpha\beta}(t_{1},t_{2};\tau_0)$:
\begin{subequations}
\label{Eq30}
\begin{equation}
\label{Eq30A}
I_{\alpha}(t;\tau_0) = 
\sum\limits_{l=-\infty}^{\infty}
e^{-il\Omega t}
{\cal J}_{\alpha,l}
e^{-\frac{1}{2}(l\Omega\tau_0)^2},
\end{equation}
\begin{equation}
\label{Eq30B}
\begin{array}{c}
{ P}_{\alpha\beta}(t_{1},t_{2};\tau_0) = 
\frac{1}{4\pi}
\int\limits_{-\infty}^{\infty} d\omega
e^{-(\omega\tau_0)^2} 
\sum\limits_{l=-\infty}^{\infty} e^{-\left(\frac{l\Omega\tau_0}{2}\right)^2}
\\
\times
e^{-i\left(l\frac{\Omega}{2} + \omega \right)t_{1}}
e^{-i\left(l\frac{\Omega}{2} - \omega \right)t_{2}} 
{\cal P}_{\alpha\beta, l}\left(l\frac{\Omega}{2} + \omega,
 l\frac{\Omega}{2} - \omega\right),
\end{array}
\end{equation}
\end{subequations}
with ${\cal J}_{\alpha,l}$ and ${\cal P}_{\alpha\beta, l}$ 
given by Eqs.(\ref{Eq23}).
Note the shift $\omega\to\omega+l\Omega/2$
we made  in Eq.(\ref{Eq30B}).

To illustrate the application of these expressions we consider 
some limiting cases.

     \subsection{Time-resolved noise of currents through the stationary conductor}

In the stationary case the scattering matrix is independent of time and 
the current spectral density, Eq.(\ref{Eq23A}), is identically zero:
 ${\cal J}_{\alpha,l}=0$. 
Therefore, the current, Eq.(\ref{Eq30A}), 
is zero, $I_{\alpha} = 0$.
In contrast, the noise due to quantum and/or thermal fluctuations is present. 
Substituting Eq.(\ref{Eq23B}) into Eq.(\ref{Eq30B}) 
and taking into account that the time-independent scattering
matrix has only a Fourier coefficient with $l=0$, 
we get:
\begin{equation}
\label{Eq31}
\begin{array}{c}
{ P}_{\alpha\beta}^{(st)}(t_{1},t_{2};\tau_0) = 
\frac{e^2}{h}
{\cal E}_{\alpha\beta}
\eta\left(\frac{t_{1}-t_{2}}{\tau_0}\right), \\
\ \\
{\cal E}_{\alpha\beta} =
2\delta_{\alpha,\beta} - |S_{\alpha\beta}|^2 - |S_{\beta\alpha}|^2, \\
\ \\
\eta(\xi) = \frac{\hbar}{2\pi\tau_0^2}
\int\limits_{0}^{\infty} dx 
e^{-i\xi x}e^{-x^2}
x \coth\left(x\frac{\hbar/\tau_0}{2k_BT}\right).
\end{array}
\end{equation}
This correlation function satisfies the conservation law
\begin{equation}
\label{Eq32}
\sum\limits_{\alpha} { P}_{\alpha\beta}^{(st)}(t_{1},t_{2};\tau_0) = 
\sum\limits_{\beta} { P}_{\alpha\beta}^{(st)}(t_{1},t_{2};\tau_0) = 0,
\end{equation}
and it depends only on the time difference as it should be 
in the stationary case 

The current correlation function ${ P}_{\alpha\beta}^{(st)}$
 is factorized into the product of two factors.
One of them, ${\cal E}_{\alpha\beta}$, 
responsible for the conservation law, Eq.(\ref{Eq32}),
depends on quantum mechanical exchange amplitudes.
This factor encodes information about the properties of a scatterer.

The second factor, $\eta(\xi)$, 
describes the temporal evolution of current fluctuations
due to intrinsic (quantum statistical) correlations in the system of Fermi particles.
\cite{Gardiner}
Only electrons originating from the same reservoir are correlated in such a way.
The electrons which originated from different reservoirs are not correlated.
Note that due to the unitarity of scattering the out-going electrons 
do not contain any additional correlations compared with incoming ones
(see, e.g., Refs.~\onlinecite{KSBK02, Wang02}).
This is why the temporal evolution of current correlations 
is described only by the function $\eta(\xi)$.

The function $\eta(\xi)$ depends crucially on the ratio 
of the measurement time interval $\tau_0$ and the temperature $T$.
We evaluate $\eta(\xi)$ in two limiting cases which we
conventionally term  "classical" and "quantum".
In the classical limit,
\begin{equation}
\label{Eq33}
\frac{h}{\tau_0} \ll k_BT,
\end{equation}
the correlation of currents is exponentially suppressed
away from a measurement time interval,
\begin{equation}
\label{Eq34}
\begin{array}{l}
\eta^{(cl)}(\xi) = 
k_BT\frac{1}{2\sqrt{\pi}\tau_0}e^{-\left(\frac{\xi}{2}\right)^2}, \\
\ \\
\xi = \frac{t_{1} - t_{2}}{\tau_0}.
\end{array}
\end{equation}
The reason is that in this limit 
the relevant current fluctuations are thermal in nature.
In the limit of $\tau_0\to 0$ the function $\eta^{(cl)}(\xi)$ 
approaches a delta-function that is characteristic for a thermal (white) noise,
$$
\lim_{\tau_0\to 0} \eta^{(cl)}\left(\frac{t_{1}-t_{2}}{\tau_0}\right) =
k_BT\delta(t_{1}-t_{2}).
$$
In contrast, in the quantum limit,
\begin{equation}
\label{Eq35}
\frac{h}{\tau_0} \gg k_BT,
\end{equation}
the function $\eta(\xi)$ is different:
\begin{equation}
\label{Eq36}
\begin{array}{c}
\eta^{(q)}(\xi) = \frac{\hbar}{2\pi\tau_0^2}\chi(\xi), \\
\ \\
\chi(\xi) = \int\limits_{0}^{\infty} dx 
\cos(\xi x) e^{-x^2}x = \left\{
\begin{array}{ll}
\frac{1}{2}, & \xi=0,\\
\ \\
-\frac{1}{\xi^2}, & \xi\gg 1.
\end{array}
\right.
\end{array}
\end{equation}
It has a long-time tail and changes sign at the crossover from
a short to a long time difference region. 
Such a behavior results form a quantum nature of fluctuations 
(with a characteristic energy $\sim h/\tau_0$) 
mainly contributing in this regime.

\begin{figure}[t]
  \vspace{0mm}
  \centerline{
   \epsfxsize 8cm
   \epsffile{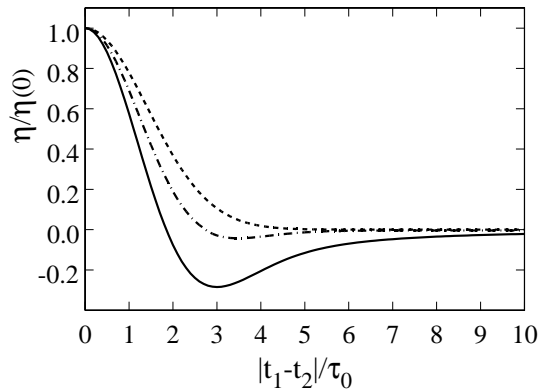}
             }
  \vspace{0mm}
  \nopagebreak
  \caption{The normalized function $\eta(\xi)/\eta(0)$ 
[see, Eq.(\ref{Eq31})]
at $\frac{k_BT}{h/\tau_0}$ = 0 (a solid line); 0.05 (a dot-dashed line);
$\infty$ (a dashed line). 
$\tau_0$ is a measurement time interval.
}
\label{fig1}
\end{figure}

In Fig.\ref{fig1} we give the function $\eta(\xi)$ 
(normalized by its value at $\xi=0$)
for several values of the parameter
$\theta = \frac{k_BT}{h/\tau_0}$.
The crossover from the quantum to classical regime occurs 
approximately at $\theta\sim 0.05$.

Note, if the time kernel is sharper than described by the Gaussian, for instance, if the currents are measured
during a time interval of duration $\tau_0$, the function $\eta(\xi)$ has qualitatively the same behavior.

Next we go to the time-dependent set-up 
and, first, consider the limit of a long measurement time.

     \subsection{Time averaged current and noise generated by a working pump}

In the limit of
\begin{equation}
\label{Eq37}
\tau_0\gg {\cal T},
\end{equation}
again only the terms with $l=0$ contribute in Eqs.(\ref{Eq30}),
since   $\Omega\tau_0\gg 1$. 
But, in contrast to the stationary case, now the scattering matrix depends on
time and, as a consequence, ac currents are generated by the pump.
At a long measurement time, Eq.(\ref{Eq37}), only the dc component
of the generated current is measured, 
$I_{\alpha}(t;\tau_0\gg{\cal T})\equiv I_{\alpha,dc}$. 
Substituting Eq.(\ref{Eq23A}) into Eq.(\ref{Eq30A}) and keeping only
the term with $l=0$, we find the well known result for 
the pumped dc current: \cite{Brouwer98}
\begin{equation}
\label{Eq38}
I_{\alpha,dc} = 
-i\frac{e}{2\pi} 
\int\limits_{0}^{{\cal T}}\frac{dt}{{\cal T}}
\left({\mathbf S}_{0}(\mu,t)
\frac{\partial{\mathbf S}_{0}^{\dagger}(\mu,t)}{\partial t} 
\right)_{\alpha\alpha},
\end{equation}
Next, calculating the current correlation function, 
Eq.(\ref{Eq30B}), with the noise power given 
by Eq.(\ref{Eq23B}) 
we take into account the following.
First, only the term with $l=0$ contributes to Eq.(\ref{Eq30B}).
Second, the relevant frequencies $\omega$ contributing to the integral 
in Eq.(\ref{Eq30B}) are of order $\tau_0^{-1}$,
see Fig.\ref{fig0}.
In the long measurement time limit they are smaller
than the pumping frequency, $\omega\ll\Omega$.
Therefore, in the sum over $q$ in Eq.(\ref{Eq23B}), 
one can drop $\omega$ in all the terms except the one with $q=0$.
In addition we suppose that the temperature is not ultra low and 
the classical limit [see, Eq.(\ref{Eq33})] holds.
Then the current correlation function is proportional to 
the zero-frequency noise power ${ P}_{\alpha\beta}$
and reads:
\begin{equation}
\label{Eq39}
\begin{array}{c}
{ P}_{\alpha\beta}(t_{1},t_{2};\tau_0\gg{\cal T}) = 
{ P}_{\alpha\beta}
\frac{1}{k_BT}\eta^{(cl)}\left(\frac{t_{1}-t_{2}}{\tau_0}\right), \\
{ P}_{\alpha\beta} = 
\frac{e^2}{h}\bigg\{ k_BT
\Big(\delta_{\alpha,\beta} + \big|{\mathit \Sigma}_{\alpha\beta}\big|^2_{0,0} 
- \big|S_{\alpha\beta}\big|_{0}^2 - \big|S_{\beta\alpha}\big|_{0}^2\Big) \\
+  \sum\limits_{q=1}^{\infty}
\frac{q\hbar\Omega}{2}\coth\left(\frac{q\hbar\Omega}{2k_BT}\right) 
\left(\big|{\mathit \Sigma}_{\alpha\beta}\big|^2_{-q,q} 
+ \big|{\mathit \Sigma}_{\alpha\beta}\big|^2_{q,-q}\right) \bigg\}.
\end{array}
\end{equation}
The zero-frequency noise power ${ P}_{\alpha\beta}$
generated by the pump was calculated in Ref.~\onlinecite{MBnoise04}.
We emphasize that here we are interested in the leading order
contribution only. 
Therefore, Eq.(\ref{Eq39}) does not contain the linear in $\Omega$ 
corrections to thermal noise investigated in Ref.~\onlinecite{MBnoise04}.

We note two consequences due to a working pump.
First, a noise survives even in the zero temperature limit (a shot-like noise).
Second, we find a modification of the thermal noise power compared 
with what we have in the stationary case. 
The nontrivial modification consists in replacing one $\delta_{\alpha,\beta}$
by the average value of a squared element of a two-particle scattering matrix
$\big|{\mathit \Sigma}_{\alpha\beta}\big|^2_{0,0}<1$.
The latter can change the sign of a linear in temperature 
contribution to the zero frequency noise power 
in the limit $k_BT\ll \hbar\Omega$,
i.e., when this contribution is only a small part of the entire noise power.

The modification of the zero frequency noise power is only
a part of the effect of an oscillating scatterer on the current correlation function
[since only the term with $l=0$ was taken into account].
To assess the whole effect 
[i.e., to take into account all the terms in Eq.(\ref{Eq30B})]
we consider the full two-time correlation function.

     \subsection{Time-resolved current and noise generated by the pump}
\label{trc}

If the measurement time interval is shorter than the pump period
then one can investigate the full time-dependent (ac) currents generated 
by the pump and their fluctuations.
In such a case, according to Eqs.(\ref{Eq30}), a number of harmonics 
$l$ contribute to the quantities of interest. 
However in each particular case it is enough to keep only
a finite number $n_{max}$ of them depending on how many of the Floquet scattering matrix elements 
${\mathbf S}_{F}(E_{n},E),n = 0, \pm 1,\dots, \pm n_{max}$
we need to correctly describe the scattering of electrons by the working pump. 
In particular, if the driving amplitude is small only the lowest side-bands matter:
$n_{max} = 1$.

Therefore, to resolve all the harmonics of a current generated, the measurement 
time interval $\tau_0$ must be small enough (but finite):
\begin{equation}
\label{Eq40}
\tau_0 \ll \frac{{\cal T}}{n_{max}}.
\end{equation}
This condition allows us to put 
$\exp\left(-\frac{1}{4}l^2\Omega^2\tau_0^2 \right)\approx 1$
in all the terms in Eqs.(\ref{Eq30}).
Substituting Eq.(\ref{Eq23A}) into Eq.(\ref{Eq30A}) and performing 
summation over $l$ we obtain a time-dependent current
$I_{\alpha}(t;\tau_0\ll {\cal T})\equiv I_{\alpha}(t)$
generated by the pump: \cite{BTP94,MBac03,BM05}
\begin{equation}
\label{Eq41}
I_{\alpha}(t) = 
-i\frac{e}{2\pi} 
\left({\mathbf S}_{0}(\mu,t)
\frac{\partial{\mathbf S}_{0}^{\dagger}(\mu,t)}{\partial t} \right)_{\alpha\alpha},
\end{equation}

We see that at enough small $\tau_{0}$ the current measured 
$I_{\alpha}(t)$ is independent of the measurement time interval.
The situation is different for the current correlation functions.
This is because with decreasing $\tau_0$ the quantum noise contribution increases, see Fig.\ref{fig0}. 

To calculate the current correlation function 
${ P}_{\alpha\beta}(t_{1},t_{2};\tau_0\ll{\cal T})\equiv
{ P}_{\alpha\beta}(t_{1},t_{2})$
we substitute Eq.(\ref{Eq23B}) into Eq.(\ref{Eq30B}).
Taking into account that the relevant frequencies
$\omega\sim\tau_0^{-1}$
are much larger than the pump frequency $\Omega$
we can decouple the integration over $\omega$ from the summation over 
$l$ and $q$. To this end we make the shift 
$\omega\to\omega - \Omega l/2$
in the first line of Eq.(\ref{Eq23B}), the shift 
$\omega\to\omega + \Omega l/2$
in the second line, and the shift
$\omega\to\omega + \Omega (l/2 - q)$
in the third line of that equation.
Then to leading order in $\Omega\tau_0$ we find:
\begin{equation}
\label{Eq42}
\begin{array}{c}
{ P}_{\alpha\beta}(t_{1},t_{2}) =
\frac{e^2}{h}
{\cal E}_{\alpha\beta}(t_{1},t_{2})
\eta\left(\frac{t_{1}-t_{2}}{\tau_0}\right), \\
\ \\
{\cal E}_{\alpha\beta}(t_{1},t_{2}) =
\delta_{\alpha,\beta}  
+ \big|{\mathit \Sigma}_{\alpha\beta}(t_{1},t_{2})\big|^2
- |S_{\alpha\beta}(t_{1})|^2 - |S_{\beta\alpha}(t_{2})|^2.
\end{array}
\end{equation}
Equation (\ref{Eq42}) resembles the result for the noise 
of a stationary scatterer, Eq.(\ref{Eq31}): 
The correlation function ${ P}_{\alpha\beta}(t_{1},t_{2})$
satisfies the conservation law, Eq.(\ref{Eq32}), 
and it contains the same factor $\eta(\xi)$ 
which originates from the correlations in a Fermi gas. 
Nevertheless the time-dependent scatterer modifies the current correlation
function considerably.
This modification concerns the factor ${\cal E}_{\alpha\beta}$ 
which depends on quantum mechanical exchange amplitudes.

To comment upon the origin of different terms in ${\cal E}_{\alpha\beta}$ 
we note that for noninteracting electrons the current correlation function can be divided 
into four statistically independent contributions originating 
from the correlation between
(i) the incoming particles,
(ii) the out-going particles, 
(iii) the particles incoming to lead $\beta$ and 
the ones out-going to lead $\alpha$, and
(iv) the particles incoming to lead $\alpha$ and 
the particles out-going to lead $\beta$.
All of them exhibit the same suppression 
-- described by the function $\eta(\xi)$ --
with increasing time difference.
However, the weight of these processes
(the four terms in ${\cal E}_{\alpha\beta}$)
differ from one another.
That is due to the difference in the relevant exchange amplitudes. 

The quantum exchange involving only incoming particles results in the term 
$\delta_{\alpha,\beta}$ reflecting the fact that electrons at different reservoirs 
are quantum-statistically independent of each other. 
This term is the same for stationary, Eq.(\ref{Eq31}), 
as well as for dynamical, Eq.(\ref{Eq42}), scattering 
because the incoming particles are assumed to be independent of
the oscillatory scatterer.
In contrast, the out-going particles are strongly affected by the pump.
As a result, the quantum exchange amplitudes involving the out-going particles 
are different for stationary and for dynamical scattering.
The correlations between incoming and out-going particles 
are proportional to the squared matrix elements of a stationary/frozen 
scattering matrix taken at the time moments when the current fluctuations are measured at the lead with the out-going particles. 
We stress that in the dynamical case, these matrix elements 
are taken at different time moments 
$t_{1}$ [in the case (iii)] and $t_{2}$ [in the case (iv)],
while in the stationary case they are time-independent.
The correlations involving only out-going particles 
changes more dramatically when the pump starts to work.
While in the stationary case these correlations are the same 
as the ones between incoming particles 
[the second $\delta_{\alpha,\beta}$ in Eq.(\ref{Eq31})],
in the dynamical case they
result in a squared matrix element of a two-particle scattering matrix 
$\big|{\mathit \Sigma}_{\alpha\beta}(t_{1},t_{2})\big|^2 $. 
For small amplitude driving, the matrix 
${\mathbf \Sigma}$ deviates from the unit matrix only a little,
${\mathit \Sigma}_{\alpha\beta}(t_{1},t_{2})\approx\delta_{\alpha,\beta}$,
while for a large amplitude driving such a deviation can be significant 
and it can reverse the sign of the entire current correlation function 
[compare Eq.(\ref{Eq31}) and Eq.(\ref{Eq42})].

To observe such a sign reversal
it is necessary to measure the current correlation function at a long
time difference, $|t_{1} - t_{2}| \sim {\cal T} \gg\tau_0$,
when the two-particle scattering matrix can differ from the unit matrix. 
On the other hand if
$\xi\equiv|t_{1} - t_{2}| /\tau_0\gg 1$ then the function $\eta(\xi)$ 
-- and hence the current correlation function ${ P}_{\alpha\beta}(t_{1},t_{2})$ --
is small. 
How small $\eta(\xi)$ depends on the ratio of the temperature and $\tau_0$.
At sufficiently low temperatures, $k_BT\ll h/\tau_0$,
the function $\eta(\xi)$ has a long-time power law tail [see, Eq.(\ref{Eq36})]
while in the opposite limit, $k_BT\gg h/\tau_0$, it is exponentially suppressed 
[see, Eq.(\ref{Eq34})] at large $\xi$.
One can consider the ratio of the dynamical and the stationary correlation 
functions and thus eliminate the common small factor $\eta(\xi)$.

In the next section we illustrate these considerations with a simple but generic example.

\section{Two terminal single channel scatterer}
  \label{2tscs}

The most general expression for the stationary scattering matrix ${\mathbf S}_{0}$ for a single-channel two-terminal conductor is (see, e.g., Ref.\onlinecite{TB99}):
\begin{equation} 
\label{Eq43} 
{\mathbf S}_{0} = e^{i\gamma}\left( 
  \begin{array}{cc} 
       \sqrt{R}e^{-i\theta}   & i\sqrt{T} e^{-i\phi}           \\ 
       i\sqrt{T}e^{i\phi}      & \sqrt{R}e^{i\theta} \\ 
  \end{array} 
\right). 
\end{equation} 
Here $R$ and $T$ are the reflection and the transmission
probability, respectively ($R+T=1$).  
We assume that the quantities entering the above equation are functions of
the external parameters  $p_j(t)$ and hence of time. 
We evaluate these quantities at the Fermi energy $\mu$.

To provide a measure of the effect of an oscillatory scatterer 
more sensitive than the current correlation function itself, 
we next introduce the current correlation coefficient
$\rho_{\alpha\beta}(t_{1},t_{2})$.

     \subsection{Current correlation coefficient}

The correlation coefficient 
(which is usual for the standard statistical analysis) calculated for currents is:
\begin{equation}
\label{Eq45}
\rho_{\alpha\beta}(t_{1},t_{2}) = 
\frac{{ P}_{\alpha\beta}(t_{1},t_{2})}{
\sqrt{\langle \Delta\hat I_{\alpha}^2(t_1)\rangle\langle \Delta\hat I_{\beta}^2(t_2)\rangle}}.
\end{equation}
This quantity is bounded, $|\rho_{\alpha\beta}|\leq 1$, and it characterizes 
the degree of correlation between the fluctuating currents measured 
at lead $\alpha$ at time moment $t_{1}$ and
at lead $\beta$ at time moment $t_{2}$:
The currents are fully correlated, anti-correlated, or non-correlated
if $\rho_{\alpha\beta} = +1, -1, 0$, respectively.
Note that this quantity can not differentiate local (i.e., classical)
and non-local (i.e., quantum-mechanical) correlations.

Substituting Eq.(\ref{Eq42}) into Eq.(\ref{Eq45}) we obtain:
\begin{equation}
\label{Eq46}
\begin{array}{c}
\rho_{\alpha\beta}(t_{1},t_{2}) = 
{\cal R}_{\alpha\beta}(t_{1},t_{2}) 
\frac{\eta\left(\frac{t_{1}-t_{2}}{\tau_0}\right)}{\eta(0)}, \\
\ \\
{\cal R}_{\alpha\beta}(t_{1},t_{2}) = 
\frac{{\cal E}_{\alpha\beta}(t_{1},t_{2})}{
\sqrt{{\cal E}_{\alpha\alpha}(t_{1},t_{1})
{\cal E}_{\beta\beta}(t_{2},t_{2})} }.
\end{array}
\end{equation}

For the two terminal stationary scatterer [see, Eq.(\ref{Eq43})]
the matrix of current correlation coefficients 
$\hat{\rho}$ (whose elements are $\rho_{\alpha\beta}$)
becomes independent of properties of the scatterer:
$$
\hat\rho^{(st)}(t_{1},t_{2}) =
\frac{\eta\left(\frac{t_{1}-t_{2}}{\tau_0} \right)}{\eta(0)} \left(
\begin{array}{cc}
 1 & -1 \\
 -1 & 1
\end{array}
\right).
$$
While for the oscillatory scatterer we get:
\begin{equation}
\label{Eq47}
\begin{array}{c}
\hat\rho(t_{1},t_{2}) =
{\cal R}(t_{1},t_{2})
\frac{\eta\left(\frac{t_{1}-t_{2}}{\tau_0} \right)}{\eta(0)} \left(
\begin{array}{cc}
 1 & -1 \\
 -1 & 1
\end{array}
\right), \\
\ \\
{\cal R}(t_{1},t_{2}) = \sqrt{T(t_{1})T(t_{2})} 
+ \sqrt{R(t_{1})R(t_{2})}\cos\big(\Delta\Theta\big), \\
\ \\
\Delta\Theta = \theta(t_{1}) - \theta(t_{2}) + \phi(t_{2}) - \phi(t_{1}).
\end{array}
\end{equation}
The correlation coefficient ${\cal R}(t_{1},t_{2})$
is sensitive to the phases of reflection and transmission coefficients.
Therefore, its behavior is nontrivial, for instance, in the case of 
a resonance-like structure which we will consider in the next section. 

Note that at coincident times $t_{1} = t_{2} = t$
the coefficient ${\cal R}(t,t) = 1$.
This means that the instant-time current correlations in the driven system 
are exactly the same as in the stationary case. 
The adiabatic pump has no effect (within the accuracy employed) on the instantaneous correlations.
In contrast, the oscillatory scatterer has a strong effect on current
correlations at different times. 
In particular, as we will see, such correlations can be suppressed 
or even their sign can be inverted compared with that of a stationary conductor.

     \subsection{Resonant transmission pump}

As a model of a resonant transmission pump we choose
a one-dimensional scatterer consisting of two delta
function barriers $V_{j}(x,t),~j = 1, 2$ oscillating with
frequency $\Omega$ and located at $x=-L/2$ and $x=L/2$:
\begin{equation}
\label{Eq48}
\begin{array}{l}
V_{1}(x,t) = \Big(V_{01} 
+ 2V_{11}\cos(\omega t + \varphi_{1})\Big)\delta\left(x+\frac{L}{2}\right), \\
\ \\
V_{2}(x,t) = \Big(V_{02} 
+ 2V_{12}\cos(\omega t + \varphi_{2})\Big)\delta\left(x-\frac{L}{2}\right), \\
\end{array}
\end{equation}
The quantities $V_{1}$ and $V_{2}$ are the pumping parameters.

The stationary scattering matrix is:
\begin{equation}
\label{Eq49}
 {\mathbf S}_{0} = \frac{e^{ikL}}{\Delta}
\left(
\begin{array}{cc}
\xi + 2\frac{p_{2}}{k}\sin(kL) & 1 \\
\ \\
1 & \xi + 2\frac{p_{1}}{k}\sin(kL)
\end{array}
\right).
\end{equation}
Here $k = \sqrt{\frac{2m}{\hbar^2}E}$; $p_j =
V_jm/\hbar^2$ (j = 1,2); $\xi$ = $(1-\Delta)e^{-ikL}$; $\Delta$ =
$1 + \frac{p_{1}p_{2}}{k^2}(e^{2ikL} -1)$ + $i\frac{p_{1}+p_{2}}{k}$. 
In numerical calculations we use the units $2m = \hbar = e = 1$. 
We take 
$L=100\pi$;
$V_{01}=V_{02}=20$;
$V_{11}=V_{12}=10$ and choose the Fermi energy $\mu$
close to the transmission resonance where the charge $\delta Q$ 
pumped for a period at 
$\varphi_{2}-\varphi_{1}=\pi/2$ 
is close to $1e$. 
We use $\mu=1.0186$ for which $\delta Q\approx 0.94e$. 

We calculate the time dependent currents 
$I_{1}(t)$, $I_{2}(t)$ and the correlation coefficient ${\cal R}(t_{1},t_{2})$ 
for several representative values of the phase difference
$\Delta\varphi \equiv \varphi_{2}-\varphi_{1} =  0,~\pi/2,~{\rm and}~ \pi$,
and consider how the peculiarities of ${\cal R}(t_{1},t_{2})$ 
relate to the peaks in $I_{1}(t)$ and $I_{2}(t)$.

For large amplitude pumps  \cite{MBnoise04} 
in the quantized pumping regime the time-averaged current and noise have the following properties:
At $\Delta\varphi = \pi/2$ the dc pumped current is maximum
and the zero-frequency noise power is minimum. 
At both $\Delta\varphi = 0$ and $\Delta\varphi = \pi$ the dc current is zero.
However the zero-frequency noise power is different.
At $\Delta\varphi = \pi$ it is close to the noise at $\Delta\varphi = \pi/2$
while at $\Delta\varphi = 0$ the noise is much larger.

          \subsubsection{$\varphi_{2} - \varphi_{1}=\frac{\pi}{2}$}

\begin{figure}[t]
  \vspace{0mm}
  \centerline{
   \epsfxsize 8cm
  \epsffile{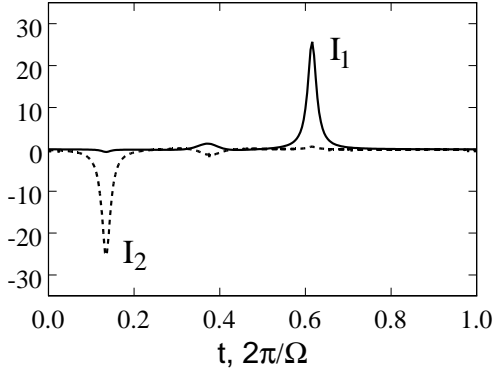}
             }
  \vspace{0mm}
  \nopagebreak
  \caption{The currents $I_1$ (in the left lead) and $I_2$ (in the right lead) 
generated by the pump are given in units of $e\Omega/(2\pi)$ as a function of time.
The parameters are: 
 $\varphi_{2} = \pi/2$;
$\varphi_{1} = 0$;
$L=100\pi$;
$V_{01}=V_{02}=20$;
$V_{11}=V_{12}=10$;
$\mu=1.018624$.
}
\label{fig2}
\end{figure}

\begin{figure}[b]
  \vspace{0mm}
  \centerline{
   \epsfxsize 8cm
  \epsffile{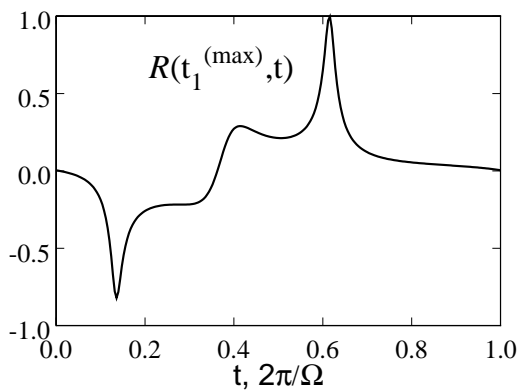}
             }
  \vspace{0mm}
  \nopagebreak
  \caption{The correlation coefficient 
${\cal R}(t_{1}^{(max)},t)$
as a function of time
for $\varphi_{2} = \pi/2$ and $\varphi_{1} = 0$. 
$t_{1}^{(max)}=0.615{\cal T}$ is a time moment when
the current $I_{1}(t)$ peaks.
Other parameters are the same as in Fig.\ref{fig2}.
}
\label{fig3}
\end{figure}

First we consider the time-dependent currents, Eq.(\ref{Eq41}), generated 
by an oscillating double-barrier pump.
In Fig.\ref{fig2} we give the currents $I_{1}(t)$ and $I_{2}(t)$
flowing in different leads.
In the quantized pumping regime the pump generates pulsed currents.
These pulses can be viewed as produced by the particles 
emitted by the pump. 
Since at different leads the currents peak at different time moments 
we conclude that the particles (an electron and a hole)
leave the pump at different time moments.

Next we consider ${\cal R}(t_{1},t_{2})$.
To get a plane graph we fix, say, the first argument  
and consider this coefficient as a function of the second argument.
We fix the first argument at that time moment $t_{1}^{(max)}$ when the
current $I_{1}$ peaks. 
From Fig.\ref{fig2} we get $t_{1}^{(max)} = 0.615{\cal T}$.

In Fig.\ref{fig3} we give the correlation coefficient ${\cal R}(t_{1}^{(max)},t)$.
At  $t=t_{1}^{(max)}=0.615{\cal T}$ the coefficient 
${\cal R}(t_{1}^{(max)},t_{1}^{(max)})=1$ as it should be.
While at the time moment 
$t=t_{2}^{(max)} = 0.135{\cal T}$ 
when the current $I_{2}$ peaks
the coefficient ${\cal R}(t_{1}^{(max)},t_{2}^{(max)})$ has a negative peak. 

Note that the correlation coefficient ${\cal R}$ can be obtained from
the measurements carried out only at lead $1$. 
One needs to measure the current and its auto-correlation function.
Nevertheless the negative peak of ${\cal R}(t_{1}^{(max)},t)$
indicates clearly when the current pulse occurs at another lead, the lead $2$. 

\begin{figure}[t]
  \vspace{0mm}
  \centerline{
   \epsfxsize 8cm
  \epsffile{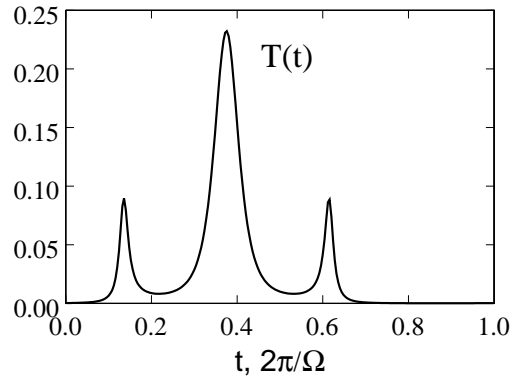}
             }
  \vspace{0mm}
  \nopagebreak
  \caption{The transmission coefficient through the pump
as a function of time
for $\varphi_{2} = \pi/2$ and $\varphi_{1} = 0$. 
Other parameters are the same as in Fig.\ref{fig2}.
}
\label{fig4}
\end{figure}

\begin{figure}[b]
  \vspace{0mm}
  \centerline{
   \epsfxsize 8cm
  \epsffile{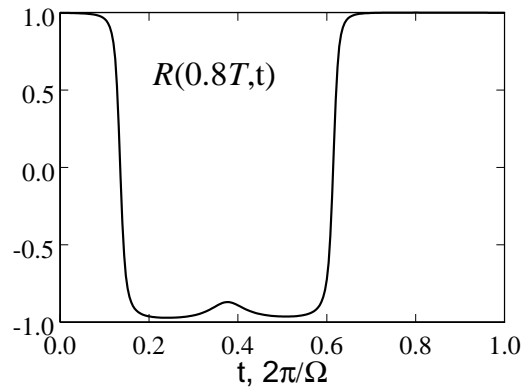}
             }
  \vspace{0mm}
  \nopagebreak
  \caption{The correlation coefficient 
${\cal R}(t_{1},t)$
as a function of time
for $\varphi_{2} = \pi/2$ and $\varphi_{1} = 0$. 
The time moment $t_{1}=0.8{\cal T}$ is taken away
from the time interval during which the current $I_{1}(t)$ peaks.
Other parameters are the same as in Fig.\ref{fig2}.
}
\label{fig5}
\end{figure}

It is instructive to compare ${\cal R}$
with the time-dependent transmission coefficient
$T(t) = \left|S_{0,12}(t) \right|^2$.
The latter is given in Fig.\ref{fig4}. 
The transmission coefficient shows small peaks at those time moments 
$t_{1}^{(max)}$ and $t_{2}^{(max)}$ when the current pulses occur. 
However it does not provide information in which lead the current pulse occurs. 
Moreover $T(t)$ shows a large peak 
when both currents $I_{1}(t)$ and $I_{2}(t)$ are small. 
Therefore, concerning the generated currents the current correlation coefficient ${\cal R}$ provides more relevant information then the transmission probability.

To characterize fully the behavior of ${\cal R}$ we 
choose a fixed time moment $t_{1}$ away from 
a current pulse interval. 
Such case is illustrated in Fig.\ref{fig5}.
The abrupt jumps of different sign 
correspond to current pulses occurring at different leads.

          \subsubsection{$\varphi_{2} - \varphi_{1}=\pi$}

\begin{figure}[t]
  \vspace{0mm}
  \centerline{
   \epsfxsize 8cm
  \epsffile{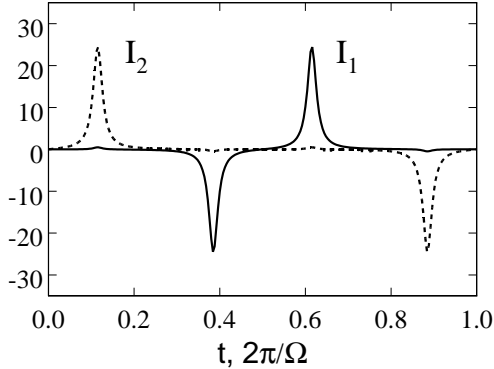}
             }
  \vspace{0mm}
  \nopagebreak
  \caption{
The currents [in units of $e\Omega/(2\pi)$] 
generated by the pump as a function of time
for $\varphi_{2} = \pi$ and $\varphi_{1} = 0$. 
Two electron-hole pairs are generated in each cycle.
Each lead receives an electron and a hole and the net pumped current is zero.
Other parameters are the same as in Fig.\ref{fig2}.
}
\label{fig6}
\end{figure}

\begin{figure}[b]
  \vspace{0mm}
  \centerline{
   \epsfxsize 8cm
  \epsffile{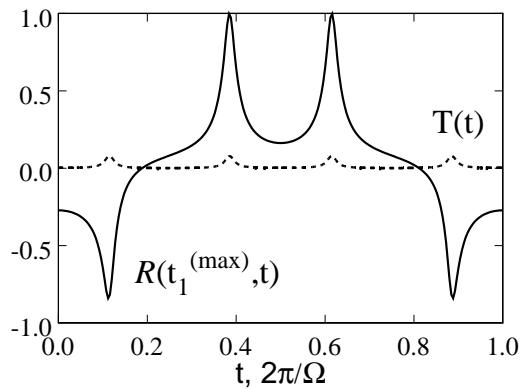}
             }
  \vspace{0mm}
  \nopagebreak
  \caption{The correlation coefficient 
${\cal R}(t_{1}^{(max)},t)$ (solid line)
and the transmission coefficient $T(t)$ (dashed line) 
as functions of time
for $\varphi_{2} = \pi$ and $\varphi_{1} = 0$. 
$t_{1}^{(max)}=0.615{\cal T}$ is a time moment
when the current $I_{1}(t)$ peaks. 
Other parameters are the same as in Fig.\ref{fig2}.
}
\label{fig7}
\end{figure}

In this case the time-dependent currents (see, Fig.\ref{fig6}), 
are similar to those we considered in the previous section.
The only difference is that for each period the pump produces two 
electron-hole pairs and pushes one electron (a positive current pulse) 
and one hole (a negative current pulse) to each lead. 
As a result the dc current is zero.

\begin{figure}[t]
  \vspace{0mm}
  \centerline{
   \epsfxsize 8cm
  \epsffile{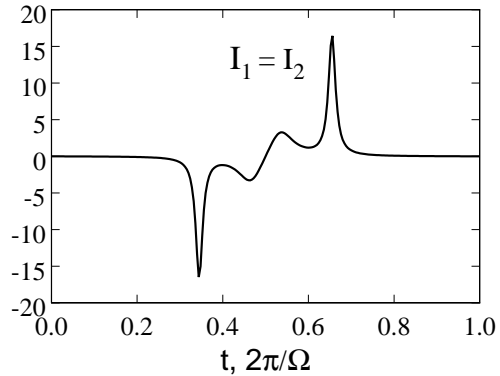}
             }
  \vspace{0mm}
  \nopagebreak
  \caption{
The currents [in units of $e\Omega/(2\pi)$] 
generated by the pump as a function of time
for $\varphi_{2} = 0$ and $\varphi_{1} = 0$. 
Other parameters are the same as in Fig.\ref{fig2}.
}
\label{fig8}
\end{figure}

In Fig.\ref{fig7} we give the correlation coefficient ${\cal R}$ 
together with the time-dependent transmission coefficient $T(t)$.
Though both ${\cal R}$ and $T(t)$ peak at those time moments
when the current $I_{1}(t)$ or $I_{2}(t)$ peaks, 
only the correlation coefficient ${\cal R}$ provides information
concerning the lead through which the current pulse occurs. 
If we fix the first argument $t_{1} = t_{1}^{(max)}$ at that time moment
when the current $I_{1}$ peaks and consider ${\cal R}(t_{1}^{(max)},t)$
as a function of the second argument $t$ then we get the following: 
At those time moments when the current $I_{1}$ peaks the coefficient
${\cal R}$ has positive peaks, while at those time moments when the current $I_{2}$ peaks the coefficient ${\cal R}$ has negative peaks.
However we should note that the correlation coefficient  ${\cal R}$
does not distinguish between the negative and positive current pulses
occurring at the same lead. 
It differentiates only the lead at which the current pulse occurs.

          \subsubsection{$\varphi_{2} - \varphi_{1}=0$}

In this regime the pump still generates approximately 
one electron-hole pair per each cycle.
The electron and hole leave the scatterer at different time moments.
However, since the potential barriers are the same, 
$V_{1}(t) = V_{2}(t)$,
the pump can push a particle into any of the leads with the same probability.
Therefore, the pump generates exactly the same currents at both leads,
see, Fig.\ref{fig8}. 
Note that such an uncertainty in the lead to which the pump pushes 
a particle results in the large zero-frequency noise mentioned already.

\begin{figure}[t]
  \vspace{0mm}
  \centerline{
   \epsfxsize 8cm
  \epsffile{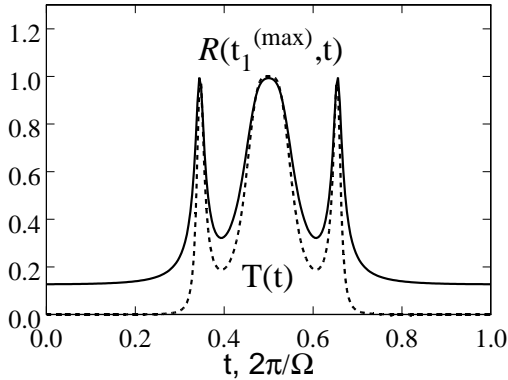}
             }
  \vspace{0mm}
  \nopagebreak
  \caption{The correlation coefficient 
${\cal R}(t_{1}^{(max)},t)$ (solid line)
and the transmission coefficient $T(t)$ (dashed line)
as functions of time
for $\varphi_{2} = 0$ and $\varphi_{1} = 0$. 
The current $I_{1}(t)$ peaks at $t_{1}^{(max)}=0.655{\cal T}$.
Other parameters are the same as in Fig.\ref{fig2}.
}
\label{fig9}
\end{figure}

Unlike the previously considered cases
now both the correlation coefficient  ${\cal R}$
and the transmission coefficient $T(t)$ 
show similar behavior, see, Fig.\ref{fig9}.
Since at both leads the current pulses occur at the same time moments,
the current correlation coefficient does not show negative peaks.
However ${\cal R}$ still shows a strong effect of an oscillatory pump on current fluctuations. 
Notice the strong suppression of correlations between the fluctuating currents at a plateau and near a peak.

\section{Conclusion}
\label{concl}

We explored the current and the noise generated by an adiabatic quantum pump in a system of noninteracting spinless electrons 
on a time-scale shorter than the pump period ${\cal T}$.
We used the Floquet scattering matrix approach which takes naturally into account 
the energy absorption and emission of electrons traversing the oscillating scatterer. 
Such many-photon processes generate ac currents 
$I_{\alpha}(\omega)$, Eq.(\ref{Eq13A}),  at frequencies 
$\omega = l\Omega$ ($l = 0, \pm 1, \dots$)
which are multiples of a pump frequency. 
The many-photon processes also correlate the fluctuating currents at frequencies $\omega$ and $\omega^{\prime}$ shifted by $l\Omega$:
$\omega = -\omega^{\prime} + l\Omega$ [see, Eq.(\ref{Eq13B})]. 

We calculated the current and the current correlation function 
averaged over a measurement time interval $\tau_0$.
In the limit of $\tau_0\gg{\cal T}$ we get the dc current, Eq.(\ref{Eq38}), and the zero frequency noise power, Eq.(\ref{Eq39}).
While in the opposite limit, $\tau_0\ll{\cal T}$, 
we find the time-dependent current $I_{\alpha}(t)$, Eq.(\ref{Eq41}),
and the two-time current correlation function 
${ P}_{\alpha\beta}(t_{1},t_{2})$, Eq.(\ref{Eq42}).
In both limits the time-dependent current and two-time correlations reveal interesting signatures of a dynamical scatterer. 
The pump is the only source of currents.
However, in general, it is only one of the sources of noise.
Other sources treated in the present paper are the thermal and equilibrium quantum fluctuations of a Fermi electron gas.
For a long measurement time, $\tau_{0}\gg {\cal T}$,
and at relatively low temperature
only the noise produced by the pump is detectable. \cite{MBnoise04}
While in a short measurement time limit, $\tau_{0}\ll {\cal T}$,
mainly the quantum fluctuations contribute to the noise measured.
Unexpectedly, in the latter regime the slow periodic pump dynamics modifies considerably the two-time current-current correlator in a long time difference limit.
The corresponding factor ${\cal E}_{\alpha\beta}(t_{1}, t_{2})$
is due to photon assisted quantum exchange 
playing a role in current fluctuations in many terminal oscillatory conductors.
The experimental investigation  of ${\cal E}_{\alpha\beta}(t_{1}, t_{2})$ is useful for several reasons:
First, it can reveal the presence of physical processes underlying 
the quantum pump effect.
Second, this quantity 
[or, alternatively, the correlation coefficient ${\cal R}_{\alpha\beta}$, 
Eq.(\ref{Eq46})]
indicates clearly the time moments at which
the current pulses generated by the pump occur.

We hope that the investigation of current-current correlations 
in a dynamically driven scatterer on a time-scale shorter than the period of a pump helps to understand deeper the nature of correlations in particle flows produced by the pump. 
\cite{SamuelssonB04,BTT05,MBen05}

\begin{acknowledgments}

This work was supported by the Swiss National Science Foundation
and the Marie Curie MCRTN-CT-2003-504574 on Fundamentals in Nanoelectronics.
M.B. acknowledges the hospitality of the Aspen Center of Physics. 

\end{acknowledgments}

\end{document}